\documentclass[aps,prl,showpacs,twocolumn,superscriptaddress]{revtex4}

\usepackage{graphicx}
\usepackage{dcolumn}
\usepackage{bm}
\usepackage{amsmath}%

\newcommand{\eepp}{\mbox{$(e,e'pp)$}}
\newcommand{\eep}{\mbox{($e,e'p$)}} 

\newcommand{\eeppn}{\mbox{$(e,e'pp)n$}}
\newcommand{\eenp}{\mbox{$(e,e'np)$}} 
\newcommand{\Het}{$^3$He}

\begin{document}


\title{Two-Nucleon Momentum Distributions Measured in $^3$He$(e,e'pp)n$}

\newcommand*{\ODU }{ Old Dominion University, Norfolk, Virginia 23529} 
\affiliation{\ODU } 

\newcommand*{\ASU }{ Arizona State University, Tempe, Arizona 85287-1504} 
\affiliation{\ASU } 

\newcommand*{\SACLAY }{ CEA-Saclay, Service de Physique Nucl\'eaire, F91191 Gif-sur-Yvette, Cedex, France} 
\affiliation{\SACLAY } 

\newcommand*{\UCLA }{ University of California at Los Angeles, Los Angeles, California  90095-1547} 
\affiliation{\UCLA } 

\newcommand*{\CMU }{ Carnegie Mellon University, Pittsburgh, Pennsylvania 15213} 
\affiliation{\CMU } 

\newcommand*{\CUA }{ Catholic University of America, Washington, D.C. 20064} 
\affiliation{\CUA } 

\newcommand*{\CNU }{ Christopher Newport University, Newport News, Virginia 23606} 
\affiliation{\CNU } 

\newcommand*{\UCONN }{ University of Connecticut, Storrs, Connecticut 06269} 
\affiliation{\UCONN } 

\newcommand*{\DUKE }{ Duke University, Durham, North Carolina 27708-0305} 
\affiliation{\DUKE } 

\newcommand*{\GBEDINBURGH }{ Edinburgh University, Edinburgh EH9 3JZ, United Kingdom} 
\affiliation{\GBEDINBURGH } 

\newcommand*{\FIU }{ Florida International University, Miami, Florida 33199} 
\affiliation{\FIU } 

\newcommand*{\FSU }{ Florida State University, Tallahassee, Florida 32306} 
\affiliation{\FSU } 

\newcommand*{\GWU }{ The George Washington University, Washington, DC 20052} 
\affiliation{\GWU } 

\newcommand*{\GBGLASGOW }{ University of Glasgow, Glasgow G12 8QQ, United Kingdom} 
\affiliation{\GBGLASGOW } 

\newcommand*{\INFNFR }{ Istituto Nazionale di Fisica Nucleare, Laboratori Nazionali di Frascati, Frascati, Italy} 
\affiliation{\INFNFR } 

\newcommand*{\INFNGE }{ Istituto Nazionale di Fisica Nucleare, Sezione di Genova, 16146 Genova, Italy} 
\affiliation{\INFNGE } 

\newcommand*{\ORSAY }{ Institut de Physique Nucleaire ORSAY, Orsay, France} 
\affiliation{\ORSAY } 

\newcommand*{\BONN }{ Institute f\"{u}r Strahlen und Kernphysik, Universit\"{a}t Bonn, Germany} 
\affiliation{\BONN } 

\newcommand*{\ITEP }{ Institute of Theoretical and Experimental Physics, Moscow, 117259, Russia} 
\affiliation{\ITEP } 

\newcommand*{\JMU }{ James Madison University, Harrisonburg, Virginia 22807} 
\affiliation{\JMU } 

\newcommand*{\KYUNGPOOK }{ Kungpook National University, Taegu 702-701, South Korea} 
\affiliation{\KYUNGPOOK } 

\newcommand*{\MIT }{ Massachusetts Institute of Technology, Cambridge, Massachusetts  02139-4307} 
\affiliation{\MIT } 

\newcommand*{\UMASS }{ University of Massachusetts, Amherst, Massachusetts  01003} 
\affiliation{\UMASS } 

\newcommand*{\UNH }{ University of New Hampshire, Durham, New Hampshire 03824-3568} 
\affiliation{\UNH } 

\newcommand*{\NSU }{ Norfolk State University, Norfolk, Virginia 23504} 
\affiliation{\NSU } 

\newcommand*{\OHIOU }{ Ohio University, Athens, Ohio  45701} 
\affiliation{\OHIOU } 

\newcommand*{\PITT }{ University of Pittsburgh, Pittsburgh, Pennsylvania 15260} 
\affiliation{\PITT } 

\newcommand*{\ROMA }{ Universita' di ROMA III, 00146 Roma, Italy} 
\affiliation{\ROMA } 

\newcommand*{\RPI }{ Rensselaer Polytechnic Institute, Troy, New York 12180-3590} 
\affiliation{\RPI } 

\newcommand*{\RICE }{ Rice University, Houston, Texas 77005-1892} 
\affiliation{\RICE } 

\newcommand*{\URICH }{ University of Richmond, Richmond, Virginia 23173} 
\affiliation{\URICH } 

\newcommand*{\SCAROLINA }{ University of South Carolina, Columbia, South Carolina 29208} 
\affiliation{\SCAROLINA } 

\newcommand*{\UTEP }{ University of Texas at El Paso, El Paso, Texas 79968} 
\affiliation{\UTEP } 

\newcommand*{\JLAB }{ Thomas Jefferson National Accelerator Facility, Newport News, Virginia 23606} 
\affiliation{\JLAB } 

\newcommand*{\UNIONC }{Union College, Schenectady, NY 12308} 
\affiliation{\UNIONC } 

\newcommand*{\VT }{ Virginia Polytechnic Institute and State University, Blacksburg, Virginia   24061-0435} 
\affiliation{\VT } 

\newcommand*{\VIRGINIA }{ University of Virginia, Charlottesville, Virginia 22901} 
\affiliation{\VIRGINIA } 

\newcommand*{\WM }{ College of William and Mary, Williamsburg, Virginia 23187-8795} 
\affiliation{\WM } 

\newcommand*{\YEREVAN }{ Yerevan Physics Institute, 375036 Yerevan, Armenia} 
\affiliation{\YEREVAN } 

\newcommand*{\deceased }{ Deceased} 


\newcommand*{\NOWNCATU }{ North Carolina Agricultural and Technical State University, Greensboro, NC 27411}

\newcommand*{\NOWGBGLASGOW }{ University of Glasgow, Glasgow G12 8QQ, United Kingdom}

\newcommand*{\NOWJLAB }{ Thomas Jefferson National Accelerator Facility, Newport News, Virginia 23606}

\newcommand*{\NOWSCAROLINA }{ University of South Carolina, Columbia, South Carolina 29208}

\newcommand*{\NOWFIU }{ Florida International University, Miami, Florida 33199}

\newcommand*{\NOWINFNFR }{ INFN, Laboratori Nazionali di Frascati, Frascati, Italy}

\newcommand*{\NOWOHIOU }{ Ohio University, Athens, Ohio  45701}

\newcommand*{\NOWCMU }{ Carnegie Mellon University, Pittsburgh, Pennsylvania 15213}

\newcommand*{\NOWINDSTRA }{ Systems Planning and Analysis, Alexandria, Virginia 22311}

\newcommand*{\NOWASU }{ Arizona State University, Tempe, Arizona 85287-1504}

\newcommand*{\NOWCISCO }{ Cisco, Washington, DC 20052}

\newcommand*{\NOWUK }{ University of Kentucky, LEXINGTON, KENTUCKY 40506}

\newcommand*{\NOWSACLAY }{ CEA-Saclay, Service de Physique Nucl\'eaire, F91191 Gif-sur-Yvette, Cedex, France}

\newcommand*{\NOWRPI }{ Rensselaer Polytechnic Institute, Troy, New York 12180-3590}

\newcommand*{\NOWDUKE }{ Duke University, Durham, North Carolina 27708-0305}

\newcommand*{\NOWUNCW }{ North Carolina}

\newcommand*{\NOWHAMPTON }{ Hampton University, Hampton, VA 23668}

\newcommand*{\NOWTulane }{ Tulane University, New Orleans, Lousiana  70118}

\newcommand*{\NOWKYUNGPOOK }{ Kungpook National University, Taegu 702-701, South Korea}

\newcommand*{\NOWCUA }{ Catholic University of America, Washington, D.C. 20064}

\newcommand*{\NOWGEORGETOWN }{ Georgetown University, Washington, DC 20057}

\newcommand*{\NOWJMU }{ James Madison University, Harrisonburg, Virginia 22807}

\newcommand*{\NOWURICH }{ University of Richmond, Richmond, Virginia 23173}

\newcommand*{\NOWCALTECH }{ California Institute of Technology, Pasadena, California 91125}

\newcommand*{\NOWMOSCOW }{ Moscow State University, General Nuclear Physics Institute, 119899 Moscow, Russia}

\newcommand*{\NOWVIRGINIA }{ University of Virginia, Charlottesville, Virginia 22901}

\newcommand*{\NOWYEREVAN }{ Yerevan Physics Institute, 375036 Yerevan, Armenia}

\newcommand*{\NOWRICE }{ Rice University, Houston, Texas 77005-1892}

\newcommand*{\NOWINFNGE }{ INFN, Sezione di Genova, 16146 Genova, Italy}

\newcommand*{\NOWBATES }{ MIT-Bates Linear Accelerator Center, Middleton, MA 01949}

\newcommand*{\NOWODU }{ Old Dominion University, Norfolk, Virginia 23529}

\newcommand*{\NOWVSU }{ Virginia State University, Petersburg,Virginia 23806}

\newcommand*{\NOWORST }{ Oregon State University, Corvallis, Oregon 97331-6507}

\newcommand*{\NOWGWU }{ The George Washington University, Washington, DC 20052}

\newcommand*{\NOWMIT }{ Massachusetts Institute of Technology, Cambridge, Massachusetts  02139-4307}

  
\author{R.A.~Niyazov}
     \affiliation{\ODU}
\author{L.B.~Weinstein}
\email[Contact Author \ ]{weinstei@physics.odu.edu}
     \affiliation{\ODU}
\author{G.~Adams}
     \affiliation{\RPI}
\author{P.~Ambrozewicz}
     \affiliation{\FIU}
\author{E.~Anciant}
     \affiliation{\SACLAY}
\author{M.~Anghinolfi}
     \affiliation{\INFNGE}
\author{B.~Asavapibhop}
     \affiliation{\UMASS}
\author{G.~Audit}
     \affiliation{\SACLAY}
\author{T.~Auger}
     \affiliation{\SACLAY}
\author{H.~Avakian}
     \affiliation{\JLAB}
     \altaffiliation{\INFNFR}
% ...

\collaboration{The CLAS Collaboration}
     \noaffiliation

\date{\today}

\begin{abstract}
We have measured the $^3$He\mbox{$(e,e'pp)n$} reaction at 2.2 GeV over a wide
kinematic range.  The kinetic energy distribution for `fast' nucleons
($p > 250$ MeV/c) peaks where two nucleons each have 20\% or less, and
the third nucleon has most of the transferred energy.  These fast $pp$
and $pn$ pairs are back-to-back with little momentum along the
three-momentum transfer, indicating that they are spectators.
Experimental and theoretical evidence indicates that we have measured
distorted two-nucleon momentum distributions by striking the {\bf third}
nucleon and detecting the spectator correlated pair.
\end{abstract}


\maketitle


The independent particle mean-field model of the nuclear wave function
is a surprisingly good approximation.  Among other successes, it
describes the shapes of the single-nucleon momentum distributions in
nuclei as measured by \eep{} nucleon knockout reactions
\cite{frullani84, kelly96, gao00}. However, discrepancies between the
measured and calculated magnitudes suggest that two-nucleon knockout
processes, especially those involving two-nucleon ($NN$) short range
correlations, are important. These short distance nucleon pairs are
primarily responsible for the high momentum components of the nuclear
wave function \cite{antonov88}.  

In addition, recent $A(e,e')$ measurements \cite{egiyan02,frankfurt93}
and theoretical calculations \cite{antonov88,forest96} indicate about
a five times higher probability per-nucleon to find an $NN$ pair with
large relative momentum and small total momentum ({\it i.e.:} in a
short range correlation) in nuclei ($A\ge 12$) than in deuterium.  We
also know that nucleons in nuclei overlap each other a significant
fraction of the time.  Taken together, these imply that we now need to
understand correlated $NN$ pairs, the next term in the mean-field
expansion of the nuclear wave function.

Unfortunately, measuring the momentum distribution of these $NN$
correlations directly is very difficult because their signals are
frequently obscured by effects such as final state interactions (FSI)
and two body currents, which include meson exchange currents (MEC) and
isobar configurations (IC) \cite{janssen00}.  To date, there have been
only a few measurements of \eepp{} or \eenp{} two nucleon knockout
from nuclei \cite{onderwater97, onderwater98, groep99, groep00}.  The
effects of correlations can only be inferred from these experiments by
comparing them to detailed calculations which include both $NN$
correlations and two body currents.  However, `exact' ({\it e.g.:}
Faddeev) calculations are only possible for light nuclei at low
energies \cite{golak95}.

The published definitions of Short Range Correlations (SRC) vary,
frequently referring to the difference between a mean field wave
function and an exact wave function.  This paper will use an
experimental definition of an SRC as an $NN$ pair with large relative
momentum and small total momentum.

This paper reports new \Het\eeppn{} results that provide a cleaner
measurement of two-nucleon momentum distributions.  Measuring these
momentum distributions will greatly aid our understanding of Short
Range Correlations.


We measured 2.261 GeV electron scattering from $^3$He, using a 100\%
duty factor beam at currents between 5 and 10 nA incident on a 4.1-cm
long liquid $^3$He target.  We detected almost all outgoing charged
particles in the Jefferson Lab CLAS (CEBAF Large Acceptance
Spectrometer), a nearly 4$\pi$ magnetic spectrometer
\cite{clas}.  These measurements were part of
the `e2' run group that took data in Spring 1999.

The CLAS uses a toroidal magnetic field and six independent sets of
drift chambers and time-of-flight scintillation counters for charged
particle identification and trajectory reconstruction.  Momentum
coverage extends down to 0.25 GeV/c for protons over a polar angular
range of $8^o < \theta < 140^o$ while spanning nearly 80\% of the
azimuth.  Electron triggers are formed from the coincidence of a gas
threshold \v Cerenkov counter and a sampling electromagnetic
calorimeter (EC).  Software fiducial cuts exclude regions
of non-uniform detector response, while  acceptance and tracking
efficiencies are estimated using GSIM, the CLAS GEANT Monte-Carlo
simulation.  

\begin{figure}[htbp]
    \includegraphics[height=2.3in]{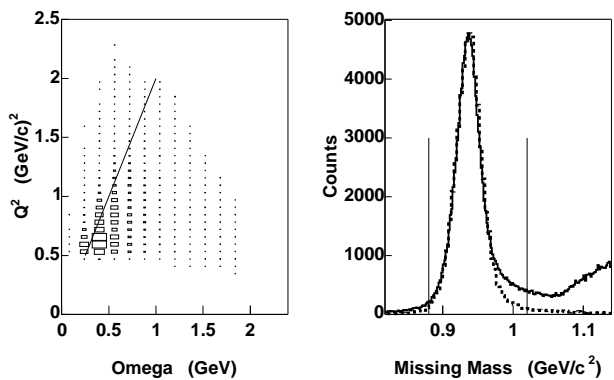} \caption{a) $Q^2$ vs
    $\omega$ for \Het\eeppn{} events.  The line shows the quasielastic
    condition $\omega = Q^2/2m_p$.  Note the large kinematic acceptance.
    b) Missing mass for \Het\eepp$X$ events.  The vertical lines
    indicate the neutron missing mass cuts.  The dashed histogram
    shows the result of a GEANT simulation of the CLAS.
    \label{fig:qomega} }
\end{figure}

We identified electrons using the total energy deposited in the EC, and
protons using time-of-flight.  We identified the neutron using missing
mass to select \Het\eeppn{} events.  We used vertex cuts to eliminate
the target walls.  Figures \ref{fig:qomega}a and b
show the electron acceptance ($Q^2 = -q_\mu q^\mu = \vec q\thinspace^2
- \omega^2$ is the square of the four-momentum transfer, $\omega$ is
the energy transfer, and $\vec q$ is the three-momentum transfer) and
undetected neutron missing mass resolution,
along with the result from a \Het\eeppn{} GSIM simulation 
that includes detector resolution but not electron radiation.
For \Het\eeppn{} events, the momentum transfer $Q^2$ is concentrated
between 0.5 and 1 (GeV/c)$^2$.  The energy transfer, $\omega$, is
concentrated slightly above but close to quasielastic kinematics
($\omega = Q^2 / 2m_p$).

We checked the data normalization by comparing \Het\eep{} cross
sections measured here and in Jefferson Lab Hall A \cite{e89044} at
the same energy and momentum transfer ($\vec q = 1.5$ GeV/c and
$\omega = 0.837$ GeV).  The ratio of our cross
sections to the Hall A cross sections was $1.00\pm0.15$, where the
error bar is  due primarily to kinematical uncertainties.

\begin{figure}[htbp]
\includegraphics[height=2.3in,width=2in]{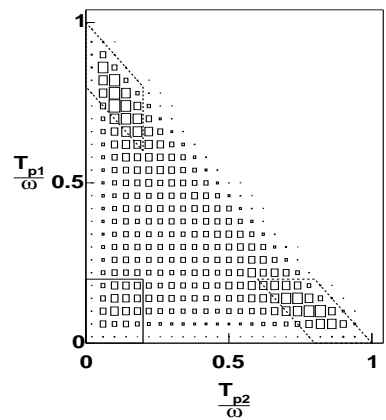} \caption{
    \Het\eeppn{} lab frame Dalitz plot.  $T_{p1}/\omega$ versus
    $T_{p2}/\omega$ for events with $p_N > 0.25$ GeV/c. \label{fig:dalitz2}}
\end{figure}

\begin{figure}[htbp]
\includegraphics[height=2.5in]{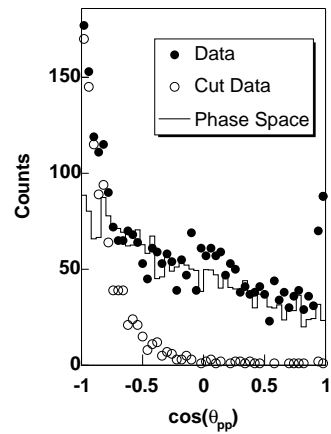} \caption{The cosine
    of the $pp$ lab frame opening angle for events with a leading
    neutron and two fast protons: $T_{p1}, T_{p2} < 0.2 * \omega$.
    Filled points show the data, open points show the data cut on
    $p^\perp_n < 300$ MeV/c, and the histogram shows the 
    phase space distribution (normalized to the data).
    \label{fig:openang2} }
\end{figure}

  
In order to understand the energy sharing in the reaction, we plotted
the kinetic energy of the first proton divided by the energy transfer
($T_{p1}/\omega$) versus that of the second proton ($T_{p2}/\omega$)
for each event (Figure \ref{fig:dalitz2}).  (Note that the assignment of
protons 1 and 2 is arbitrary.)  Since the threshold for proton
detection is $p_p = 250$ MeV/c, we also cut on neutron momentum $p_n
\ge 250$ MeV/c.  There are three peaks at the three corners of the
plot, corresponding to events where two nucleons each have less than
20\% of the energy transfer and the third `leading' nucleon has the
remainder.  We call the two nucleons `fast' because $p > 250$ MeV/c is
larger than the average nucleon bound-state momentum.  We cut on these peaks,
as indicated by the lines in Figure \ref{fig:dalitz2}.  The solid lines
indicate the `leading $n$, fast $pp$ pair' cut and the dashed lines
indicate the `leading $p$, fast $pn$ pair' cut.

Then we looked at the opening angle of the two fast nucleons.  Figure
\ref{fig:openang2} shows the opening angle for fast $pp$ pairs with a
leading neutron (the opening angle distribution of fast $pn$ pairs for
events with a leading proton is almost identical).  Note the large
peak at 180$^o$ ($\cos\theta_{NN} \approx -1$).  
The peak is not due to the cuts, since we do not see it in a 
simulation which assumes three-body absorption of the
virtual photon followed by phase space decay \cite{pdg-phasespace}. It
is also not due to the CLAS acceptance since we see it for both fast
$pp$ and fast $pn$ pairs.  This back-to-back peak is a very strong
indication of correlated $NN$ pairs.

Now that we have identified correlated pairs, we want to study them.
In order to reduce the effects of final state rescattering, we cut on
the perpendicular component (relative to $\vec q$\thinspace) of the
leading nucleon's momentum, $p^\perp < 0.3$ GeV/c.  The resulting fast
$NN$ pair opening angle distribution is almost entirely back-to-back
(see Figure \ref{fig:openang2}).  These fast nucleons are distributed
almost isotropically in angle (after correcting for the CLAS
acceptance).  The pair average total momentum parallel to $\vec q$
($<p_{tot}^\parallel> \sim 0.05$ GeV/c) is also much smaller than the
average $q$ ($<q> \sim 1$ GeV/c).

Both of these indicate that the  paired nucleons are predominantly
spectators and that their measured  momentum distributions reflect the
pair's initial momentum distribution in the nucleus.

\begin{figure}[htbp]
    \includegraphics[width=3.5in,height=4.2in,angle=0]{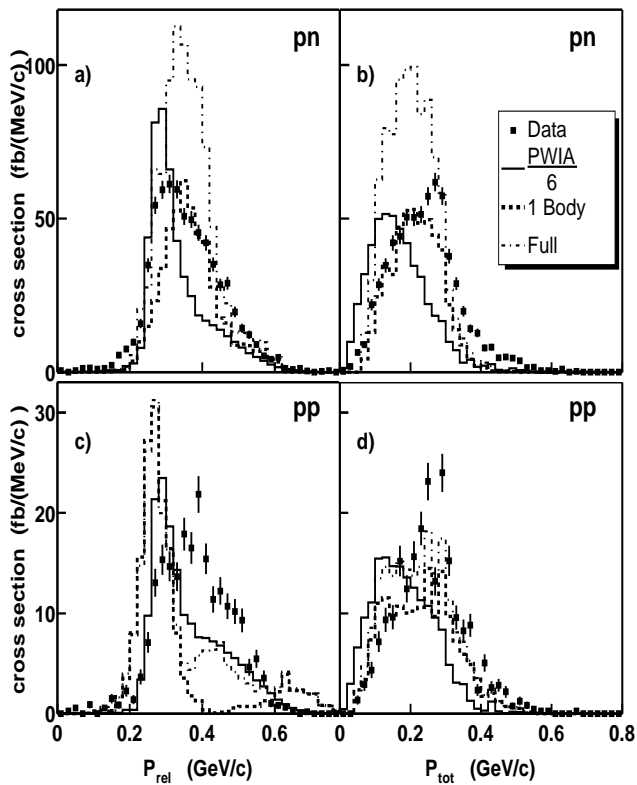}
    \caption{a) Lab frame cross section vs. relative momentum of the
    fast $pn$ pair.  Points show the data, solid histogram shows the
    PWIA calculation reduced by a factor of 6 \cite{misakpc}, thick dashed
    histogram shows Laget's one-body calculation
    \cite{laget88,audit97,laget87}, thin dot-dashed histogram shows Laget's
    full calculation; b) the same for the total momentum; c) and d)
    the same for fast $pp$ pairs. \label{fig:prelptot} }
\end{figure}

The resulting relative $\vec p_{rel} = (\vec p_1 - \vec p_2)/2$ and
total $\vec p_{tot} = \vec p_1 + \vec p_2$ momentum distributions of
the $pn$ and $pp$ pairs are shown in Figure \ref{fig:prelptot}.  Since
the $NN$ pairs are spectators, all quantities and cross sections are
given in the lab frame.  The cross sections are integrated over the
experimental acceptance.  Radiative and tracking efficiency
corrections have been applied \cite{RustamPhd}.  
The overall normalization uncertainty is 15\%.  

\begin{figure}[htbp]
    \includegraphics[height=1.5in]{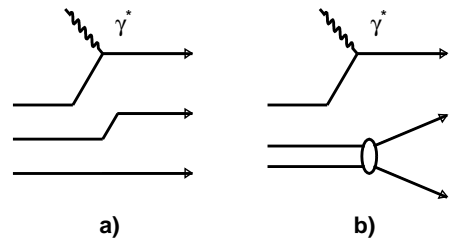}
    \caption{Feynman diagrams for a) Plane Wave Impulse Approximation
    and b) pair distortion.
\label{fig:feynmann} }
\end{figure}

The relative momentum distribution rises rapidly starting at about
0.25 GeV/c (limited by the minimum nucleon momenta of 0.25 GeV/c),
peaks at about 0.35 GeV/c, and has a tail extending to about 0.7 GeV/c.
The total momentum distribution rises rapidly from 0, peaks at about
0.25 GeV/c, and falls rapidly.  The momentum distributions have an
upper limit determined by the cut $T_{fast} < 0.2*\omega$.  Note that
these distributions are very similar for both $pp$ and $pn$ pairs.

\begin{table}[hbtp]
\begin{tabular}{|l|r|r|}\hline
Cross Section 	& \multicolumn{2}{c|}{} 	 \\ 
(pb)	& $pp$ 		& $pn$  	\\ \hline         
Data          		& 4.4$\pm$0.1	& 13.4$\pm$0.2		\\ \hline 
Laget 1-body 		& 3.3		& 9.9		\\ \hline
Laget Full		& 4.2		& 18.6		\\ \hline
PWIA 			& 20.7		& 60.5		\\ \hline         
PWIA / Data		& 4.8		& 4.5		\\ \hline         
\end{tabular}
\caption{Cross sections integrated over the CLAS acceptance.  The
normalization uncertainty (systematic uncertainty) of the data is
15\%.  The calculations are described in the text.}
\label{tab:sigma}
\end{table}

We also compared our data to a Plane Wave Impulse Approximation (PWIA)
calculation (see Figure \ref{fig:feynmann}a) by Sargsian
\cite{misakpc} that uses an exact $^3$He wave function \cite{nogga00}
and the De Forest `cc1' single nucleon current \cite{deForest83}.  We
generated events in phase space, weighted them by the PWIA cross
section, and applied the same cuts
as with the actual data.  The results are reasonably close
(considering the simplicity of the model) to the data except for a
scale factor.  The data distributions have similar shapes, including
the virtual photon distribution, the kinetic energy distributions, and
the fast pair opening angles.  The relative momentum distributions are
similar but have a different detailed shape (see the solid histograms in
Figure \ref{fig:prelptot}).  The PWIA total momentum distribution peaks
significantly below the data.  These discrepancies will be discussed
below.

Table \ref{tab:sigma} shows the integrated cross sections.  The PWIA
cross section is on average about five times larger than the data for
both $pp$ and $pn$ pairs.  Note that the ratio of $pn$ to $pp$ cross
sections is approximately the same for data (3.0) and for PWIA (2.9)
indicating the importance of single particle knockout in the reaction
mechanism.

Exact calculations by W. Gl\"ockle {\it et al.} \cite{glockle01} at
much lower momentum transfer and $p_{tot}=0$ looked at the effects of
different reaction mechanisms.  They found that neither MEC nor
rescattering of the leading nucleon  had an effect, and that the
continuum state interaction of the outgoing $NN$ pair (`pair distortion'
-- diagram b of Figure \ref{fig:feynmann}) decreased the cross section by a factor
of approximately 10 relative to the PWIA result.  Calculations by
C. Ciofi degli Atti and L. Kaptari also found that pair distortion
significantly decreased the cross section \cite{cda02}.

Calculations by Laget (described in detail below) also showed these
effects.  His calculation further showed that pair distortion reduces
the PWIA cross section for $s$-wave $NN$ pairs much more than for
$p$-wave pairs, effectively shifting both the $p_{rel}$ and $p_{tot}$
peaks to higher momentum.  Laget's one-body $p_{tot}$ distribution
peaks at about 250 MeV/c, much larger than the PWIA $p_{tot}$ peak and
in better agreement with the data (see the thick dashed curve in
Figure \ref{fig:prelptot}b and d).

Thus, these calculations suggest that the factor of five difference
between the data and the PWIA calculation (Figure \ref{fig:feynmann}a)
is due to the continuum state interaction of the outgoing $NN$ pair
(pair distortion -- Figure \ref{fig:feynmann}b).  That plus the rough
similarity between the data and the PWIA calculation indicates that we
may have measured two-nucleon momentum distributions by striking the
third nucleon and observing the spectator correlated pair.

We also compared our data to a full calculation using a diagrammatic
approach by Laget \cite{laget88,audit97,laget87}, integrated over the
CLAS acceptance \cite{jmlpc}. This calculation includes one-, two- and
three-body amplitudes.  The one-body amplitudes include diagrams with
two spectator nucleons including direct knockout (Figure
\ref{fig:feynmann}a) plus continuum state interaction of the spectator
$NN$ pair (pair distortion (Figure \ref{fig:feynmann}b)). The two-body
amplitudes include diagrams with one spectator nucleon including FSI
between the struck nucleon and one other plus two-body MEC and IC
\cite{laget87}. The three-body amplitudes include diagrams with no
spectator nucleons including three-body MEC and IC \cite{laget88}.
The calculation uses the dominant $s$- and $p$-waves for the $T=1$
pairs and $s$- and $d$-waves for the $T=0$ pairs that are then
coupled to the third nucleon in the bound state wave function.  The
model made absolute predictions and was not adjusted to fit the data.

The one-body calculations describe the $pn$ pairs well, both
qualitatively and quantitatively (see Figures \ref{fig:prelptot}a and b).
However, the full calculation overestimates the data by about 60\%.
The calculation describes $p_{rel}$ for $pp$ pairs badly but $p_{tot}$
well (see Figures \ref{fig:prelptot}c and d). The failure is due possibly
to the truncation of the wave function to only the lower angular
momentum states.  Note that Laget predicts three-body effects to be
much larger for events with a leading proton and a fast $pn$ pair than
for events with a leading neutron and a fast $pp$ pair.  We do not see
this difference in the data.

Comparison of the results of Laget's calculations with the data shows
that (1) the continuum state interaction of the outgoing $NN$ pair
decreases the cross section significantly relative to the PWIA result,
and by supressing the $s$-wave, shifts the peak to larger momenta, (2)
two-body currents (MEC and IC) plus rescattering of the leading
nucleon contribute less than 5\% of the cross section, and (3)
three-body currents contribute about 20\% of the $pp$ and 50\% of the
$pn$ cross section, but do not improve agreement with the data.

These results  reinforce the conclusions we drew from the data that
we are measuring the high momentum part of the distorted  $NN$
momentum distribution.  Note however, that since
two-body currents do not contribute, the only other possible contributions
are due to three-body currents, also a subject of great interest.

Detailed calculations with exact wave functions are clearly needed in
order to quantitatively relate the measured distorted $NN$ momentum
distributions to Short Range Correlations in the nucleus.


To summarize, we have measured the \Het\eeppn{} reaction at 2.2  GeV over a
wide kinematic range.  The kinetic energy distribution for `fast'
nucleons ($p > 250$ MeV/c) peaks where two nucleons each have 20\% or
less and the third or `leading' nucleon carries most of the
transferred energy.  These fast nucleon pairs (both $pp$ and $pn$) are
back-to-back, almost isotropic, and carry very little momentum along $\vec
q$, indicating that they are predominantly spectators.

PWIA calculations reproduce the observed $pp$ to $pn$ cross section
ratio, indicating the importance of single-nucleon knockout mechanisms.
Calculations by Laget with many different diagrams and a truncated
bound state wave function  predict that leading-nucleon FSI and two-body
exchange currents are negligible, and continuum-state interactions of
the spectator pair reduce the cross section significantly.  However, the
predicted three-body exchange current contributions of about 20\% for
$pp$ pairs and 50\% for $pn$ pairs do not improve agreement with the
data.  

Thus experimental and theoretical evidence indicates
that we have measured distorted  $NN$ momentum distributions in
\Het\eeppn{} by striking the {\bf third} nucleon and detecting the
spectator correlated pair.

\begin{acknowledgments}
We  acknowledge the outstanding efforts of the staff of
the Accelerator Division and the Physics Division (especially the CLAS target
group) at Jefferson Lab that made this experiment possible.

This work was supported in part by the Italian Istituto Nazionale di Fisica
Nucleare, the French Centre National de la Recherche Scientifique, the
French Commissariat \`{a} l'Energie Atomique, the U.S. Department of
Energy, the National Science Foundation, Emmy Noether grant from the
Deutsche Forschungsgemeinschaft and the Korean Science and
Engineering Foundation.  The Southeastern Universities Research
Association (SURA) operates the Thomas Jefferson National Accelerator
Facility for the United States Department of Energy under contract
DE-AC05-84ER40150.

\end{acknowledgments}

\bibliography{eep,3he,clas}

\begin{thebibliography}{26}
\expandafter\ifx\csname natexlab\endcsname\relax\def\natexlab#1{#1}\fi
\expandafter\ifx\csname bibnamefont\endcsname\relax
  \def\bibnamefont#1{#1}\fi
\expandafter\ifx\csname bibfnamefont\endcsname\relax
  \def\bibfnamefont#1{#1}\fi
\expandafter\ifx\csname citenamefont\endcsname\relax
  \def\citenamefont#1{#1}\fi
\expandafter\ifx\csname url\endcsname\relax
  \def\url#1{\texttt{#1}}\fi
\expandafter\ifx\csname urlprefix\endcsname\relax\def\urlprefix{URL }\fi
\providecommand{\bibinfo}[2]{#2}
\providecommand{\eprint}[2][]{\url{#2}}

\bibitem[{\citenamefont{Frullani and Mougey}(1984)}]{frullani84}
\bibinfo{author}{\bibfnamefont{S.}~\bibnamefont{Frullani}} \bibnamefont{and}
  \bibinfo{author}{\bibfnamefont{J.}~\bibnamefont{Mougey}},
  \bibinfo{journal}{Adv. Nucl. Phys.} \textbf{\bibinfo{volume}{14}},
  \bibinfo{pages}{1} (\bibinfo{year}{1984}).

\bibitem[{\citenamefont{Kelly}(1996)}]{kelly96}
\bibinfo{author}{\bibfnamefont{J.}~\bibnamefont{Kelly}}, \bibinfo{journal}{Adv.
  Nucl. Phys.} \textbf{\bibinfo{volume}{23}}, \bibinfo{pages}{75}
  (\bibinfo{year}{1996}).

\bibitem[{\citenamefont{Gao et~al.}(2000)}]{gao00}
\bibinfo{author}{\bibfnamefont{J.}~\bibnamefont{Gao}} \bibnamefont{et~al.},
  \bibinfo{journal}{Phys. Rev. Lett.} \textbf{\bibinfo{volume}{84}},
  \bibinfo{pages}{3265} (\bibinfo{year}{2000}).

\bibitem[{\citenamefont{Antonov et~al.}(1988)\citenamefont{Antonov, Hodgson,
  and Petkov}}]{antonov88}
\bibinfo{author}{\bibfnamefont{A.}~\bibnamefont{Antonov}},
  \bibinfo{author}{\bibfnamefont{P.}~\bibnamefont{Hodgson}}, \bibnamefont{and}
  \bibinfo{author}{\bibfnamefont{I.}~\bibnamefont{Petkov}},
  \emph{\bibinfo{title}{Nucleon Momentum and Density Distributions in Nuclei}}
  (\bibinfo{publisher}{Clarendon Press}, \bibinfo{year}{1988}).

\bibitem[{\citenamefont{Egiyan et~al.}(2003)}]{egiyan02}
\bibinfo{author}{\bibfnamefont{K.}~\bibnamefont{Egiyan}} \bibnamefont{et~al.},
  \bibinfo{journal}{Phys. Rev. C, in press}  (\bibinfo{year}{2003}),
  \eprint[http://arXiv.org/abs]{nucl-ex/0301008}.

\bibitem[{\citenamefont{Frankfurt et~al.}(1993)\citenamefont{Frankfurt,
  Strikman, Day, and Sargsyan}}]{frankfurt93}
\bibinfo{author}{\bibfnamefont{L.}~\bibnamefont{Frankfurt}},
  \bibinfo{author}{\bibfnamefont{M.}~\bibnamefont{Strikman}},
  \bibinfo{author}{\bibfnamefont{D.}~\bibnamefont{Day}}, \bibnamefont{and}
  \bibinfo{author}{\bibfnamefont{M.}~\bibnamefont{Sargsyan}},
  \bibinfo{journal}{Phys. Rev. C} \textbf{\bibinfo{volume}{48}},
  \bibinfo{pages}{2451} (\bibinfo{year}{1993}).

\bibitem[{\citenamefont{Forest et~al.}(1996)}]{forest96}
\bibinfo{author}{\bibfnamefont{J.}~\bibnamefont{Forest}} \bibnamefont{et~al.},
  \bibinfo{journal}{Phys. Rev. C} \textbf{\bibinfo{volume}{54}},
  \bibinfo{pages}{646} (\bibinfo{year}{1996}).

\bibitem[{\citenamefont{Janssen et~al.}(2000)}]{janssen00}
\bibinfo{author}{\bibfnamefont{S.}~\bibnamefont{Janssen}} \bibnamefont{et~al.},
  \bibinfo{journal}{Nucl. Phys.} \textbf{\bibinfo{volume}{A672}},
  \bibinfo{pages}{285} (\bibinfo{year}{2000}).

\bibitem[{\citenamefont{Onderwater et~al.}(1997)}]{onderwater97}
\bibinfo{author}{\bibfnamefont{G.}~\bibnamefont{Onderwater}}
  \bibnamefont{et~al.}, \bibinfo{journal}{Phys. Rev. Lett.}
  \textbf{\bibinfo{volume}{78}}, \bibinfo{pages}{4893} (\bibinfo{year}{1997}).

\bibitem[{\citenamefont{Onderwater et~al.}(1998)}]{onderwater98}
\bibinfo{author}{\bibfnamefont{G.}~\bibnamefont{Onderwater}}
  \bibnamefont{et~al.}, \bibinfo{journal}{Phys. Rev. Lett.}
  \textbf{\bibinfo{volume}{81}}, \bibinfo{pages}{2213} (\bibinfo{year}{1998}).

\bibitem[{\citenamefont{Groep et~al.}(1999)}]{groep99}
\bibinfo{author}{\bibfnamefont{D.}~\bibnamefont{Groep}} \bibnamefont{et~al.},
  \bibinfo{journal}{Phys. Rev. Lett.} \textbf{\bibinfo{volume}{83}},
  \bibinfo{pages}{5443} (\bibinfo{year}{1999}).

\bibitem[{\citenamefont{Groep et~al.}(2000)}]{groep00}
\bibinfo{author}{\bibfnamefont{D.}~\bibnamefont{Groep}} \bibnamefont{et~al.},
  \bibinfo{journal}{Phys. Rev. C} \textbf{\bibinfo{volume}{63}},
  \bibinfo{pages}{014005} (\bibinfo{year}{2000}).

\bibitem[{\citenamefont{Golak et~al.}(1995)}]{golak95}
\bibinfo{author}{\bibfnamefont{J.}~\bibnamefont{Golak}} \bibnamefont{et~al.},
  \bibinfo{journal}{Phys. Rev. C} \textbf{\bibinfo{volume}{51}},
  \bibinfo{pages}{1638} (\bibinfo{year}{1995}).

\bibitem[{\citenamefont{Mecking et~al.}(2003)}]{clas}
\bibinfo{author}{\bibfnamefont{B.}~\bibnamefont{Mecking}} \bibnamefont{et~al.},
  \bibinfo{journal}{Nucl. Inst. and Meth.} \textbf{\bibinfo{volume}{A503}},
  \bibinfo{pages}{513} (\bibinfo{year}{2003}).

\bibitem[{\citenamefont{Higinbotham}(2002)}]{e89044}
\bibinfo{author}{\bibfnamefont{D.}~\bibnamefont{Higinbotham}}, in
  \emph{\bibinfo{booktitle}{Proceedings of the International Symposium on
  Electromagnetic Interactions in Nuclear and Hadronic Physics}}, edited by
  \bibinfo{editor}{\bibfnamefont{M.}~\bibnamefont{Fujiwara}} \bibnamefont{and}
  \bibinfo{editor}{\bibfnamefont{T.}~\bibnamefont{Shima}}
  (\bibinfo{publisher}{World Scientific}, \bibinfo{year}{2002}), p.
  \bibinfo{pages}{291}.

\bibitem[{\citenamefont{Hagiwara et~al.}(2002)}]{pdg-phasespace}
\bibinfo{author}{\bibfnamefont{K.}~\bibnamefont{Hagiwara}}
  \bibnamefont{et~al.}, \bibinfo{journal}{Phys. Rev. D}
  \textbf{\bibinfo{volume}{66}}, \bibinfo{pages}{010001}
  (\bibinfo{year}{2002}).

\bibitem[{\citenamefont{Niyazov}(2003)}]{RustamPhd}
\bibinfo{author}{\bibfnamefont{R.}~\bibnamefont{Niyazov}}, Ph.D. thesis,
  \bibinfo{school}{Old Dominion University} (\bibinfo{year}{2003}).

\bibitem[{\citenamefont{Sargsian}()}]{misakpc}
\bibinfo{author}{\bibfnamefont{M.}~\bibnamefont{Sargsian}},
  \emph{\bibinfo{title}{Private communication}}.

\bibitem[{\citenamefont{Laget}(1988)}]{laget88}
\bibinfo{author}{\bibfnamefont{J.}~\bibnamefont{Laget}}, \bibinfo{journal}{J.
  Phys. G} \textbf{\bibinfo{volume}{14}}, \bibinfo{pages}{1445}
  (\bibinfo{year}{1988}).

\bibitem[{\citenamefont{Audit et~al.}(1997)}]{audit97}
\bibinfo{author}{\bibfnamefont{G.}~\bibnamefont{Audit}} \bibnamefont{et~al.},
  \bibinfo{journal}{Nucl. Phys.} \textbf{\bibinfo{volume}{A614}},
  \bibinfo{pages}{461} (\bibinfo{year}{1997}).

\bibitem[{\citenamefont{Laget}(1987)}]{laget87}
\bibinfo{author}{\bibfnamefont{J.}~\bibnamefont{Laget}},
  \bibinfo{journal}{Phys. Rev. C} \textbf{\bibinfo{volume}{35}},
  \bibinfo{pages}{832} (\bibinfo{year}{1987}).

\bibitem[{\citenamefont{Nogga et~al.}(2001)\citenamefont{Nogga, Kamada, and
  Gl{\" o}ckle}}]{nogga00}
\bibinfo{author}{\bibfnamefont{A.}~\bibnamefont{Nogga}},
  \bibinfo{author}{\bibfnamefont{H.}~\bibnamefont{Kamada}}, \bibnamefont{and}
  \bibinfo{author}{\bibfnamefont{W.}~\bibnamefont{Gl{\" o}ckle}},
  \bibinfo{journal}{Nucl. Phys.} \textbf{\bibinfo{volume}{A689}},
  \bibinfo{pages}{357} (\bibinfo{year}{2001}),
  \eprint[http://arXiv.org/abs]{nucl-th/0010005}.

\bibitem[{\citenamefont{De~Forest}(1983)}]{deForest83}
\bibinfo{author}{\bibfnamefont{T.}~\bibnamefont{De~Forest}},
  \bibinfo{journal}{Nucl. Phys.} \textbf{\bibinfo{volume}{A392}},
  \bibinfo{pages}{232} (\bibinfo{year}{1983}).

\bibitem[{\citenamefont{Gl{\" o}ckle et~al.}(2001)}]{glockle01}
\bibinfo{author}{\bibfnamefont{W.}~\bibnamefont{Gl{\" o}ckle}}
  \bibnamefont{et~al.}, \bibinfo{journal}{Acta Phys. Polon.}
  \textbf{\bibinfo{volume}{B32}}, \bibinfo{pages}{3053} (\bibinfo{year}{2001}),
  \eprint{nucl-th/0109070}.

\bibitem[{\citenamefont{Ciofi~degli Atti and Kaptari}(2002)}]{cda02}
\bibinfo{author}{\bibfnamefont{C.}~\bibnamefont{Ciofi~degli Atti}}
  \bibnamefont{and} \bibinfo{author}{\bibfnamefont{L.}~\bibnamefont{Kaptari}},
  \bibinfo{journal}{Phys. Rev. C} \textbf{\bibinfo{volume}{66}},
  \bibinfo{pages}{044004} (\bibinfo{year}{2002}),
  \eprint[http://arXiv.org/abs]{nucl-th/0203041}.

\bibitem[{\citenamefont{Audit and Laget}()}]{jmlpc}
\bibinfo{author}{\bibfnamefont{G.}~\bibnamefont{Audit}} \bibnamefont{and}
  \bibinfo{author}{\bibfnamefont{J.-M.} \bibnamefont{Laget}},
  \emph{\bibinfo{title}{Private communication}}.

\end{thebibliography}

\end{document}